\documentclass[aps,twocolumn,showpacs,floatfix]{revtex4}

\usepackage{amsfonts}
\usepackage{amsmath}
\usepackage{amssymb}
\usepackage{graphicx}
\usepackage{color}
\usepackage{verbatim}
\usepackage{dsfont}
\usepackage{hyperref}
\usepackage{xfrac}
\usepackage[normalem]{ulem}

\begin{document}

\title{Hyperfine level structure in nitrogen-vacancy centers near the ground-state level anticrossing}
\author{Marcis Auzinsh$^1$}
\author{Andris Berzins$^1$}
\author{Dmitry Budker$^{2,3}$}
\author{Laima Busaite$^1$}
\email{laima.busaite@lu.lv}
\author{Ruvin Ferber$^1$}
\author{Florian Gahbauer$^1$}
\author{Reinis Lazda$^1$}
\email{reinis.lazda@lu.lv}
\author{Arne Wickenbrock$^2$}
\author{Huijie Zheng$^2,$}
\email{zheng@uni-mainz.de}

\affiliation{$^1$Laser Centre, University of Latvia}
\affiliation{$^2$Helmholtz Institute Mainz, Johannes Gutenberg University}
\affiliation{$^3$Department of Physics, University of California at Berkeley, USA}

%76.30.Mi Color centers and other defects
%77.22.Ej Polarization and depolarization
%76.70.Hb Optically detected magnetic resonance (ODMR)
%75.10.Dg Crystal-field theory and spin Hamiltonians (see also 71.70.Ch Crystal and ligand fields)

\begin{abstract}
Energy levels of nitrogen-vacancy centers in diamond were investigated using optically detected magnetic-resonance spectroscopy near the electronic ground-state level anticrossing (GSLAC) at an axial magnetic field around 102.4~mT in diamond samples with a nitrogen concentration of 1~ppm and 200~ppm. By applying radiowaves in the frequency ranges from 0 to 40 MHz and from 5.6 to 5.9 GHz, we observed transitions that involve energy levels mixed by the hyperfine interaction. We developed a theoretical model that describes the level mixing, transition energies, and transition strengths between the ground-state sublevels, including the coupling to the nuclear spin of the NV center\textquotesingle s $^{14}$N and $^{13}$C atoms. The calculations were combined with the experimental results by fitting the ODMR spectral lines based on a theoretical model, which yielded information about the polarization of nuclear spins. This study is important for the optimization of experimental conditions in GSLAC-based applications, e.g., microwave-free magnetometry and microwave-free nuclear-magnetic-resonance probes.
\end{abstract}

\maketitle
\section{Introduction}

Nitrogen-vacancy (NV) color centers in diamond crystals currently are used in a broad range of applications. They serve as qubits~\cite{Popkin1} or quantum-memory elements~\cite{Heshami1} for quantum computers, or probes for various physical quantities like magnetic field~\cite{Zheng1,Wickenbrock1}, electric field ~\cite{Dolde1,Wrachtrup1}, strain~\cite{Tamarat1,Manson1}, rotation~\cite{Bud4,Wood2,Ash1} or temperature~\cite{Clevenson1}. They can also be used to detect the properties of electronic and nuclear spins on the surface or in the interior of a diamond crystal~\cite{Mamin1,Staud1,Loretz1,Muller1}, such as substitutional nitrogen (P1) centers~\cite{Wood1,Wickenbrock1, Bud2}, $^{13}{\rm C}$ atoms~\cite{Wood1}, and cross-relaxation with other point defects in the diamond lattice~\cite{Wang1}. The presence and properties of other spin centers can be ascertained by measuring longitudinal ($T_1$) or transverse ($T_2$) relaxation times of the polarization of the NV centers' ground-state electron spins.

For these applications it is crucially important to know in detail the energy level structure of the NV center, including its hyperfine structure, which arises from the interaction of the electron spin with the nuclear spin of the $^{14}$N atom which is a part of the NV center. One electronic magnetic sublevel split by the hyperfine interaction interacts with another electronic magnetic sublevel split by the hyperfine interaction. Near the magnetic field values at which magnetic sublevels cross or have avoided crossings (e.g., GSLAC), this interaction leads to strong hyperfine level mixing and alters the transition probabilities that involve these mixed levels.

The interaction of NV centers with nearby $^{13}\mathrm{C}$ atoms and their nuclear spin polarization has been studied using electron spin resonance at low magnetic fields and near the excited state level anti-crossing at 512 G~\cite{Dre1}. It has been shown by studying optically detected magnetic resonance (ODMR)~\cite{Negyedi1} signals that $^{13}\mathrm{C}$ nuclei adjacent to the vacancy can be polarized using only optical methods at specific values of the magnetic field near the GSLAC. The hyperfine manifold and level anticrossings of the NV center with the nuclear spin of $^{14}\mathrm{N}$ and $^{15}\mathrm{N}$ has been studied in the presence of a magnetic field of several tens of gauss transverse to the NV axis~\cite{Clevenson2}. All-optical methods have been used to study the hyperfine structure induced by the interaction of NV centres with their nitrogen atoms for the case of $^{14}\mathrm{N}$ and $^{15}\mathrm{N}$~\cite{Broa1}, whereas the dynamic nuclear polarization of $^{15}\mathrm{N}$ as a function of magnetic field was modelled up to the GSLAC~\cite{Iva1}.

In this study, we used the ODMR method to investigate the ground state $m_s= 0 \longrightarrow m_s=\pm 1$ electron spin transitions and to study the hyperfine level structure of NV-center ensembles in the vicinity of the GSLAC. We calculated the level structure of these electron-spin states and the microwave-field-induced transition strengths between these levels. Then we used a parameter-optimization procedure to fit the experimentally measured curves with the results of the theoretical calculation. This fitting procedure yielded information about the degree of nuclear polarization of the $^{14}\mathrm{N}$ spin in the vicinity of the GSLAC. The theoretical model gave the relative intensities of the transitions and, by adding coupling to $^{13}\mathrm{C}$, made it possible to describe additional transitions in the measured signals. 
The model applied a Monte Carlo approach to include the interaction with $^{13}\mathrm{C}$ nuclei (which make up 1.1\% of the carbon nuclei) for those lattice positions out to a distance of 5~\AA \,  that have significant coupling.

\begin{figure*}%[tb]
  \begin{center}
    \includegraphics[width=0.95\textwidth]{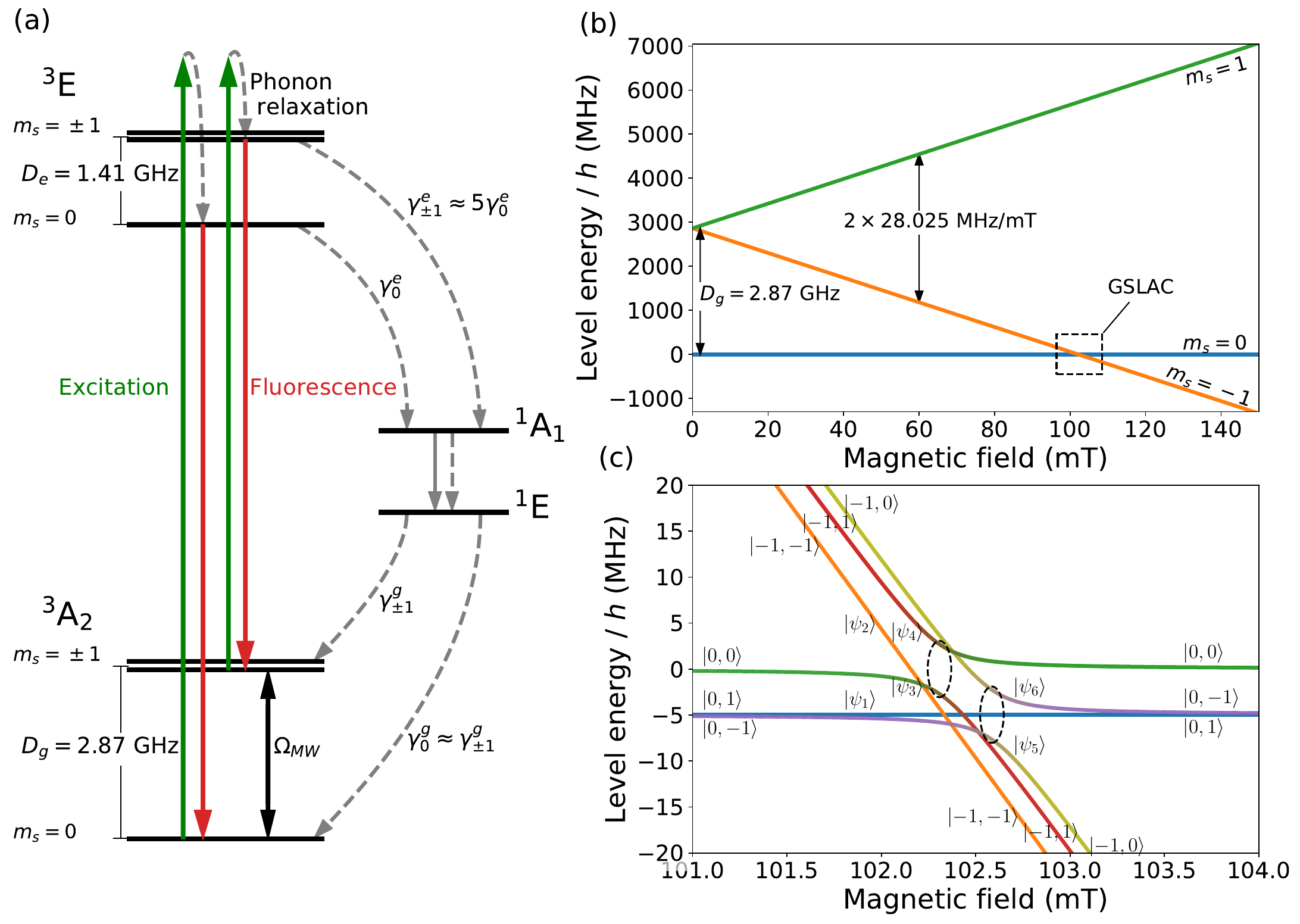}
  \end{center}
  \caption{(a) Level scheme for an NV center in diamond where $m_S$ is the electron spin projection quantum number, $D_g$ and $D_e$ are the ground- and excited-state zero-magnetic-field splittings, $\Omega_{MW}$ is the MW Rabi frequency, $\gamma_0^g$ and $\gamma_{\pm 1}^g$ are the relaxation rates from the singlet state $^1$E to the triplet ground state $^3$A$_2$, $\gamma_0^e$ and $\gamma_{\pm 1}^e$ are the relaxation rates from the triplet excited state $^3$E to the singlet state $^1$A$_1$. (b) Levels of the NV center's electron-spin magnetic sublevels in the ground state. (c) Hyperfine level ($|m_S,m_I\rangle$) anticrossing in the vicinity of the GSLAC. The degree of mixing near the GSLAC (denoted by the dashed ellipses) is indicated by the relative admixture of the colors in each curve; the lines corresponding to unmixed states do not change color.}
  \label{fig_gross}
\end{figure*}

\section{Hyperfine level anticrossing in NV centers in diamond}
\label{Sec_model}

The NV center is composed of a substitutional nitrogen atom and an adjacent vacancy. It exists in different charged states NV$^0$ and NV$^-$. In this work we focus on the energy levels of the NV$^-$ and refrain from writing out the charge state. The NV center has an electron spin $S=1$ in the ground state. There are both triplet and singlet excited states as shown in Fig.~\ref{fig_gross}(a).

In the absence of an external magnetic field the NV center has a splitting between the ground-state magnetic sublevels of the electron spin $m_S = 0$ and $m_S=\pm1$ due to the spin-spin interaction~\cite{Doherty1}. This zero-field splitting corresponds to 2.87 GHz in the $^3$A$_2$ ground state and about 1.41 GHz in the $^3$E excited state. If an external magnetic field $\bf B$ is applied along the NV axis, the magnetic sublevels of the electron spin acquire additional energy equal to

\begin{equation}
E_{m_S }= \gamma_{e} B m_S,
\end{equation}
where $\gamma_{e}=28.025$ GHz/T is the electron gyromagnetic ratio, $m_S$ is the magnetic quantum number of the electron spin, and $B$ is the magnetic field strength.

In addition to the electronic states, interactions with nearby nuclear spins must be considered. In all cases, the nucleus of the nitrogen atom associated with the NV center interacts with the NV electron spin. The vast majority (99.6\%) of these nitrogen nuclei are $^{14}{\rm N}$ whose nuclear spin is $I=1$. Furthermore, although $^{12}{\rm C}$ nuclei have zero nuclear spin, some ($\sim 1.1$\%) of the nearby carbon nuclei are $^{13}{\rm C}$ with nuclear spin $I= \frac{1}{2}$.
There are interactions between the NV center and nearby P1 centers but we will not consider this interaction here.

In this approximation the Hamiltonian for the NV center in its ground state can be written as \cite{Wood1}:
\begin{equation}
 \hat{H} = \hat{H}_{el}+\hat{H}_{14N}+\hat{H}_{NV+14N}+\hat{H}_{13C}+\hat{H}_{NV+13C}, 
 \label{eq1}
\end{equation}
where $\hat{H}_{el}=D_g \hat S_z^2 + \gamma_e \mathbf{B} \cdot \mathbf{\hat S}$ describes the ground state of the NV center with electron spin $\mathbf{S}$, gyromagnetic ratio $\gamma_{e}$, and zero-field splitting $D_g=2.87$ GHz; $\hat{H}_{14N}=Q \hat I_z^2 - \gamma_{14N} \mathbf{B} \cdot \mathbf{\hat I}$ describes the $^{14}{\rm N}$ nucleus with spin $\mathbf{I}$, electric quadrupole interaction parameter $Q = - 4.96$ MHz, and gyromagnetic ratio $\gamma_{14N}=3.077$ MHz/T; $\hat{H}_{13C}=\Sigma_j\gamma_{13C} \mathbf{B} \cdot \mathbf{\hat J}_j$ describes $^{13}{\rm C}$ nuclei with nuclear spin $\mathbf{J}_j$ and gyromagnetic ratio $\gamma_{13C}=10.704$ MHz/T in the external magnetic field; $\hat{H}_{NV+14N}=\mathbf{\hat S} \cdot \hat{A} \cdot \mathbf{\hat I}$ describes the hyperfine interaction of the NV center with the $^{14}$N nucleus via the diagonal hyperfine interaction tensor $\hat{A}$; and $\hat{H}_{NV+13C}$ describes the interaction of the $^{13}{\rm C}$ nucleus and the NV center. The last term requires special consideration, since the strength of the interaction depends on the distance between the NV center and the $^{13}{\rm C}$ nucleus or nuclei and their relative orientation. In this section we will describe the calculation of the energy levels and interaction strengths without the $\hat{H}_{NV+13C}$ or $\hat{H}_{13C}$ terms. In Sec.~\ref{sec:C13} we will discuss the effect of adding in the interactions with $^{13}$C.
 
The matrix $\hat{A}$ is a diagonal hyperfine-interaction tensor between the electron spin $\bf S$ of the NV center and nuclear spin $\bf I$ of the $^{14}$N nucleus that belongs to the NV center,

\begin{equation}
\hat{A} =
\left(
\begin{array}{ccc}
A_{\perp} & 0 & 0  \\ 
0 & A_{\perp} & 0 \\
0 & 0 & A_{\parallel}  
\end{array} 
\right) ,
\label{AN14}
\end{equation}
where the hyperfine-interaction parameters are $A_\parallel = - 2.14$ MHz, 
$A_\perp = -2.70$ MHz.
The values of the constants are taken from~\cite{Wood1} and references therein.

A crossing between $m_S=0$ and $m_S=-1$ occurs when the Zeeman splitting compensates the zero-field splitting at a magnetic field value of $D_g/\gamma_{e}=$102.4 mT [see Fig.~\ref{fig_gross}(b)]. Owing to the hyperfine interaction each of the electron-spin substates are split into hyperfine components. Some of the hyperfine components exhibit avoided crossings [see Fig. \ref{fig_gross}(c)].

The contribution of the $m_S=+1$ sublevel and its hyperfine components can be plausibly ignored when calculating the eigenvalues and eigenvectors of the $m_S=0$ and $m_S=-1$ hyperfine components near the GSLAC since the $m_S=+1$ sublevel is separated from the other two by an energy corresponding to 5740 MHz. After calculating the eigenvalues and eigenvectors of the $m_S=0$ and $m_S=-1$ submanifold, we added the states $\psi_7$, $\psi_8$, and $\psi_9$ [see~\eqref{eigenfunctions}], which belong the the $m_S=+1$ manifold, unmixed in this approximation, to have a full set of levels. Thus, we used a ``truncated" Hamiltonian that includes only the electron spin sublevels with $m_S = 0$ and $m_S=-1$ and the nuclear spin sublevels with $m_I = 0, \pm 1$, to obtain approximate analytical solutions for the energy levels and wave functions of the hyperfine states. The energies $E_i$ of these components are

\begin{widetext}
\begin{subequations}
\begin{align}
E_1 & =  Q  , \\
E_2 & =  D_g + Q + A_\parallel - \gamma_e B, \\
E_3 & =  \frac{1}{2} \left(D_g + Q  - A_\parallel - \gamma_e B - \sqrt{4 A_\perp^2 + (D_g + Q  - A_\parallel - \gamma_e B)^2 } \right) , \\
E_4 & =  \frac{1}{2} \left(D_g + Q  - A_\parallel - \gamma_e B + \sqrt{4 A_\perp^2 + (D_g + Q  - A_\parallel - \gamma_e B)^2} \right)  ,  \\
E_5 & =  \frac{1}{2} \left(D_g + Q - \gamma_e B  - \sqrt{4 A_\perp^2 + (D_g - Q - \gamma_e B)^2}\right) , \\
E_6 & =  \frac{1}{2} \left(D_g + Q - \gamma_e B  + \sqrt{4 A_\perp^2 + (D_g - Q - \gamma_e B)^2}\right), \\
E_7 &= D_g + Q + A_\parallel + \gamma_e B, \\
E_8 &=  D_g + Q - A_\parallel + \gamma_e B, \\
E_9 &= D_g + \gamma_e B.
\end{align}\label{energies}
\end{subequations}
\end{widetext}

The wave functions can be written in the uncoupled basis $|m_S, m_I\rangle$ as follows:

\begin{widetext}
\begin{subequations}
\begin{align}
\vert \psi_1 \rangle & = \vert 0, 1 \rangle , \\
\vert \psi_2 \rangle & = \vert -1, -1 \rangle , \\
\vert \psi_3 \rangle & = \frac{ 1 }{ \vert \alpha_1 \vert } \vert -1, 1 \rangle - \frac{ 1 }{ \vert \alpha_1 \vert } \left( \kappa_1 + \sqrt{\kappa_1^2 + 1} \right) \vert 0,0 \rangle , \\
\vert \psi_4 \rangle & = \frac{ 1 }{ \vert \alpha_1 \vert } \vert -1, 1 \rangle - \frac{ 1 }{ \vert \alpha_1 \vert } \left( \kappa_1 - \sqrt{\kappa_1^2 + 1} \right) |0,0 \rangle , \\
\vert \psi_5 \rangle & = \frac{ 1 }{ \vert \alpha_2 \vert } \vert -1, 0 \rangle - \frac{ 1 }{ \vert \alpha_2 \vert } \left( \kappa_2 + \sqrt{\kappa_2^2 + 1} \right) \vert 0,-1 \rangle , \\
\vert \psi_6 \rangle & = \frac{ 1 }{ \vert \alpha_2 \vert } \vert -1, 0 \rangle - \frac{ 1 }{ \vert \alpha_2 \vert  } \left( \kappa_2 - \sqrt{\kappa_2^2 + 1} \right) \vert 0,-1 \rangle , \\
\vert \psi_7 \rangle & = \vert 1, 1 \rangle , \\
\vert \psi_8 \rangle & = \vert 1, -1 \rangle ,  \\
\vert \psi_9 \rangle & = \vert 1, 0 \rangle ,
\end{align}\label{eigenfunctions}
\end{subequations}

where
\begin{subequations}
\begin{align}
\kappa_1 & = \frac{D_g + Q - A_\parallel - \gamma_e B}{2 A_\perp} ,\\
\kappa_2 & = \frac{D_g - Q - \gamma_e B}{2 A_\perp}
\end{align}\label{eigenfunctionsa}
\end{subequations}

and
\begin{equation}
\vert \alpha_{1,2}| = \sqrt{\left( \kappa_{1,2} + \sqrt{ \kappa_{1,2}^2+1 } \right)^2+1}.
\label{eigenfunctionsb}
\end{equation}
\end{widetext}

In the magnetic field some hyperfine levels become mixed. Although the magnetic quantum numbers $m_S$ and $m_I$ cease to be good quantum numbers as a result of this mixing, their sum $m_S+m_I$ still is preserved [see Eq.~\eqref{eigenfunctions} (c)--(f)]. Only states with equal $m_S+m_I$ interact [see the interactions denoted in Fig.~\ref{fig_gross} (c)]. 
In particular, the states $\psi_1$ and $\psi_2$ remain unmixed even in the strong magnetic field that corresponds to the GSLAC. The other four states form two pairs of mixed states. One pair consists of states $\psi_3$ and $\psi_4$, the other of states $\psi_5$ and $\psi_6$. This information about the state mixing will be important when we analyze which transitions are allowed and which are forbidden when the magnetic field value is close to $102.4$~mT.

Magnetic-dipole transitions between various states can be driven with an applied radio-frequency magnetic field. The selection rules for these transitions are $\Delta m_S = \pm 1$ and $\Delta m_I = 0$ ~\cite{Macq1,Sann1,Donghun1}.

Having calculated the energy levels and wave functions in \eqref{energies} and \eqref{eigenfunctions}, respectively, we now want to describe transitions between different states. We will consider only transitions that change the electron spin state $\mathbf{S}$ in~\eqref{eq1}. In order to describe electron spin transitions we must add to the Hamiltonian a term to describe the interaction with the microwave (MW) field. This term can be constructed starting from the raising and lowering operators:
\begin{equation}
\label{S-hat}
\hat{S}_{\pm}=\hat{S}_x \pm i\hat{S}_y,
\end{equation}
where $\hat{S}_x \pm i\hat{S}_y$ are the spin operators for $S=1$.
However, we need a $9\times 9$ matrix to describe the raising and lowering of the electron spin $S=1$ in a system that contains also a nuclear spin $I=1$. This operator $S'_{\pm}$ can be obtained by taking the outer product with the three-dimensional identity matrix $\mathds{1}(3)$ and folding in the matrix of wave functions $\Psi$ whose columns are the ground-state eigenvectors $\vert \phi_i \rangle$.
Then the interaction term can be written as
\begin{equation}
\hat{H}^{(m)}_{INT} = \frac{\Omega_{MW}}{2}\left( \hat{S}'_+ +\hat{S}'_- \right) \; , \\
%\begin{split}
%&=\Omega \left( {\begin{array}{ccc} 
%	d'_{00} & d'_{0-1} & d'_{01}\\
%	d'_{-10} & d'_{-1-1} & d'_{-11}\\
%	d'_{10} & d'_{1-1} & d'_{11}\\
%	\end{array}} \right) \otimes \mathds{1}(3) .
%	\end{array}} \right) \otimes \mathds{1}(3) ,
%\end{split}
\end{equation}
%where the $d_{ij}$ are dipole transition matrix elements. 
where $\Omega_{MW}$ is the Rabi frequency of the microwave radiation at the given transition frequency.
We thus obtain a $9\times 9$ block-diagonal matrix of magnetic dipole transition elements $m_{ij}$.
From these transition matrix elements $m_{ij}$ the transition probabilities $p'_{ij}$ between levels $i$ and $j$ can be obtained: 
\begin{equation}
\label{transition_strengths}
p'_{ij}=m'_{ij}\cdot m'_{ji} \; .
\end{equation}
To obtain actual transition intensities $t_{ij}$ we multiply the transition probabilities by a term that takes into account the actual populations of each level, which depends on the polarization of $^{14}\mathrm{N}$.
We describe this polarization  using the concept of ``spin temperature" $\beta$~\cite{DDD_2004}, which is defined as
\begin{equation}
\label{spin_temp}
P_{m_I}(\beta)=\frac{e^{-m_I\cdot \beta}}{e^{-\beta}} \; .
\end{equation}
Now the observed transition strengths will be the product of the transition probability and a term that depends on the populations of the levels involved in the transition. We define a matrix $\mathds{M}$ whose dimensions match the matrix of transition probabilities in~\eqref{transition_strengths}, and whose elements $M_{ij}=\sqrt{P_{m_i} P_{m_f}}$~\cite{DeLimaBernardo2017}, where $m_i$ is the nuclear spin of the initial state in the transition and $m_f$ is the nuclear spin of the final state. Now we multiply the terms in \eqref{transition_strengths} and $\mathds{M}$ to obtain the transition intensities:
\begin{equation}
    \label{eq12}
    t_{ij}=p'_{ij}\cdot \mathds{M}_{ij}  \; .
\end{equation}
We can then construct the calculated ODMR spectrum in which the integral under the ODMR peak that corresponds to the energy difference between levels $i$ and $j$ is proportional to  transition strength $t_{ij}$.
%Now we rewrite the magnetic dipole Hamiltonian in the new basis:
%\begin{equation}
%\hat{H}'^{(m)}=\frac{1}{2}\left( \hat{S}'_{+}+\hat{S}'_{-} \right) = 
%\left( {\begin{array}{ccc} 
%	d'_{00} & d'_{0-1} & d'_{01}\\
%	d'_{-10} & d'_{-1-1} & d'_{-11}\\
%	d'_{10} & d'_{1-1} & d'_{11}\\
%	\end{array}} \right) \otimes \mathds{1}(3) .
%	\end{array}} \right) \otimes \mathds{M} .
%    \label{transition_strengths}
%\end{equation}
%And the transition probability is proportional to
%\begin{equation}
%p'_{ij}=d'_{ij}\cdot d'_{ji} .
%\end{equation}

\section{Experimental set-up}
\label{Sec_exp}
We measured ODMR signals from ensembles of NV centers in two different samples. One sample was produced by chemical vapor deposition with a nitrogen concentration around 1~ppm (low-density sample). The other sample was a dense high-pressure, high-temperature (HPHT)
crystal with a relatively high concentration of nitrogen
of around 200 ppm (high-density sample). The measurements with the low-density sample were performed at the Johannes Gutenberg-University in Mainz, whereas the measurements with the high-density sample were performed at the Laser Centre of the University of Latvia in Riga. The NV centers were irradiated with green 532~nm light from a Nd:YAG laser (Coherent Verdi) and optically polarized to the $m_S=0$ state while the luminescence from the $^3E$ state was monitored [see the transition diagram in Fig.~\ref{fig_gross}(a)]. Following the ODMR method, a microwave field was applied to induce transitions between the ground-state sublevels. The NV centers' electrons were continuously pumped to the $m_S=0$ state. When a MW field is on resonance with a transition from an $m_S=0$ hyperfine component to an $m_S=\pm 1$ hyperfine component, the fluorescence intensity decreases.
\begin{figure}%[tb]
  \begin{center}
    \includegraphics[width=0.45\textwidth]{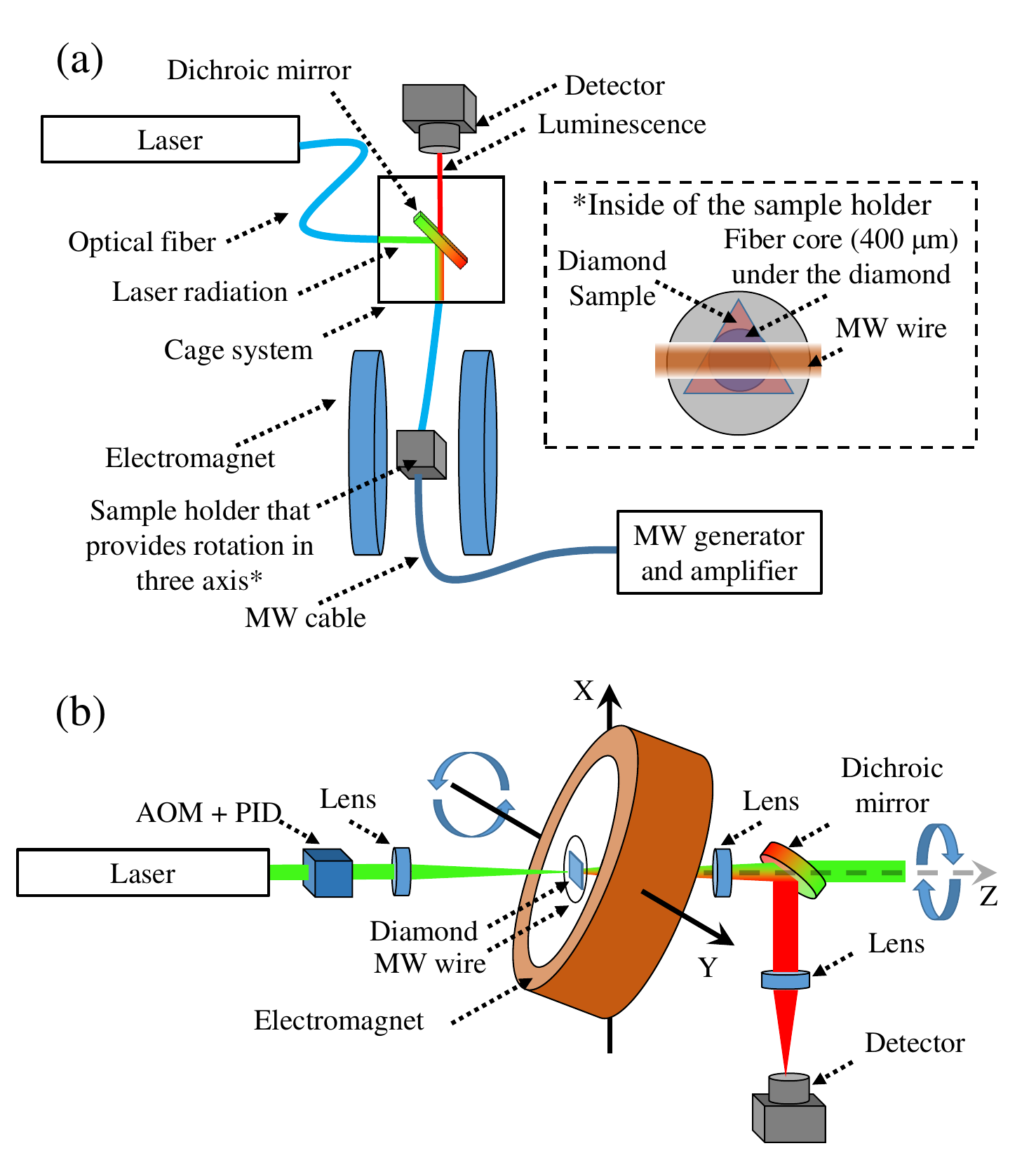}
  \end{center}
  \caption{Experimental setup. (a) High-density sample. The laser light was coupled into an optical fiber that led it to a cage-system cross, in which a dichroic mirror (Thorlabs DMLP567R) reflects the green light, which, in turn, was coupled into another fiber, which led to the sample. A small portion of the green light that passed through the dichroic mirror was used to monitor the laser power. The fluorescence from the sample, which was glued to the end of the fiber, was collected with the same fiber, and, after passing through the dichroic mirror and a long-pass filter (Thorlabs FEL0600), it was focused onto a photodiode with an amplifier (Thorlabs PDA36A-EC). (b) Low-density sample. Laser light was focused on the sample with a lens, fluorescence was collected and measured with a photodetector (Thorlabs APD410A/M). The diamond sample is located at the center of the magnet bore. The rotation axes of the sample and the electromagnet are depicted with blue arrows.}
  \label{setup}
\end{figure}

Figure~\ref{setup}(a) shows the experimental setup used for the high-density sample in Riga.
The magnetic field was produced by a custom-built magnet initially designed for electron paramagnetic resonance (EPR) experiments. It consists of two 19 cm diameter iron poles with a length of 13 cm each, separated by a 5.5 cm air gap. This magnet could provide a highly homogeneous field (0.0002 mT over the sensing volume, estimated by a simulation software COMSOL modeling). The diamond sample under investigation was held in place using a nonmagnetic holder (custom-made by STANDA), which provided three axes of rotation to align the NV axis with the applied magnetic field. Light with a wavelength of 532 nm (Coherent Verdi Nd:YAG) was delivered to the sample via an optical fiber with a core diameter of 400 micrometers (numerical aperture of 0:39). The same fiber was used to collect red  fluorescence light, which was separated from the residual green reflections by a dichroic mirror and a long-pass filter (Thorlabs DMLP567R and FEL0600) and focused onto an amplified photodiode (Thorlabs PDA36A-EC). The signals were recorded and averaged on a digital oscilloscope (Agilent DSO5014A or Yokogawa DL6154) or a DAQ card (Measurement Computing USB-1408FS).

Figure~\ref{setup}(b) shows the experimental setup used for the low-density sample in Mainz. 
A custom-made electromagnet was used with  200 turns wound on a water-cooled copper mount. The electromagnet produced a field of 2.9~mT/A and could achieve magnetic fields up to 103.5 mT. The diamond could be rotated around the z-axis (NV axis). Moreover, the electromagnet could be moved with a computer-controlled 3D translation stage (Thorlabs PT3-Z8) and a rotation stage (Thorlabs NR360S, y-axis). In this way, all degrees of freedom for centering the diamond in the magnet and aligning the NV axis to the magnetic field were available. An accousto-optic modulator (AOM) in combination with a photodiode and a proportional-integral-derivative (PID) controller served to stabilize the laser intensity.

In both setups, the microwave field was generated and amplified using two sets of devices depending on the required frequency range. In Riga, for low frequencies a TTi TG5011 generator (0.001 mHz to 50 MHz) and for high frequencies a function generator (SRS SG386) with a power amplifier (Minicircuits ZVE-3W-83+) provided up to +30dBm. In Mainz, an SRS SG386 was used as a function generator over the entire range, in conjunction with power amplifiers. At high frequencies, an RFLU PA0706GDRF amplifier (Lambda) was used. It was replaced at low frequencies with a (Minicircuits ZHL-32A+) amplifier.

%An SRS SG386 microwave generator provided microwave signals over the entire range in conjunction with power amplifiers (Lambda RFLUPA0706GD at high frequencies; Minicircuits ZHL-32A+ at low frequencies.)

\section{Experimental results and analysis}
\label{sec:analysis}
\subsection{ODMR signals for the \texorpdfstring{$\vert m_S=0 \rangle \longrightarrow \vert m_S=+1 \rangle$}{|mS=0>-->mS=+1} transition}\label{sec:ODMR_hf}

\begin{figure*}%[tb]
  \begin{center}
    \includegraphics[width=\textwidth]{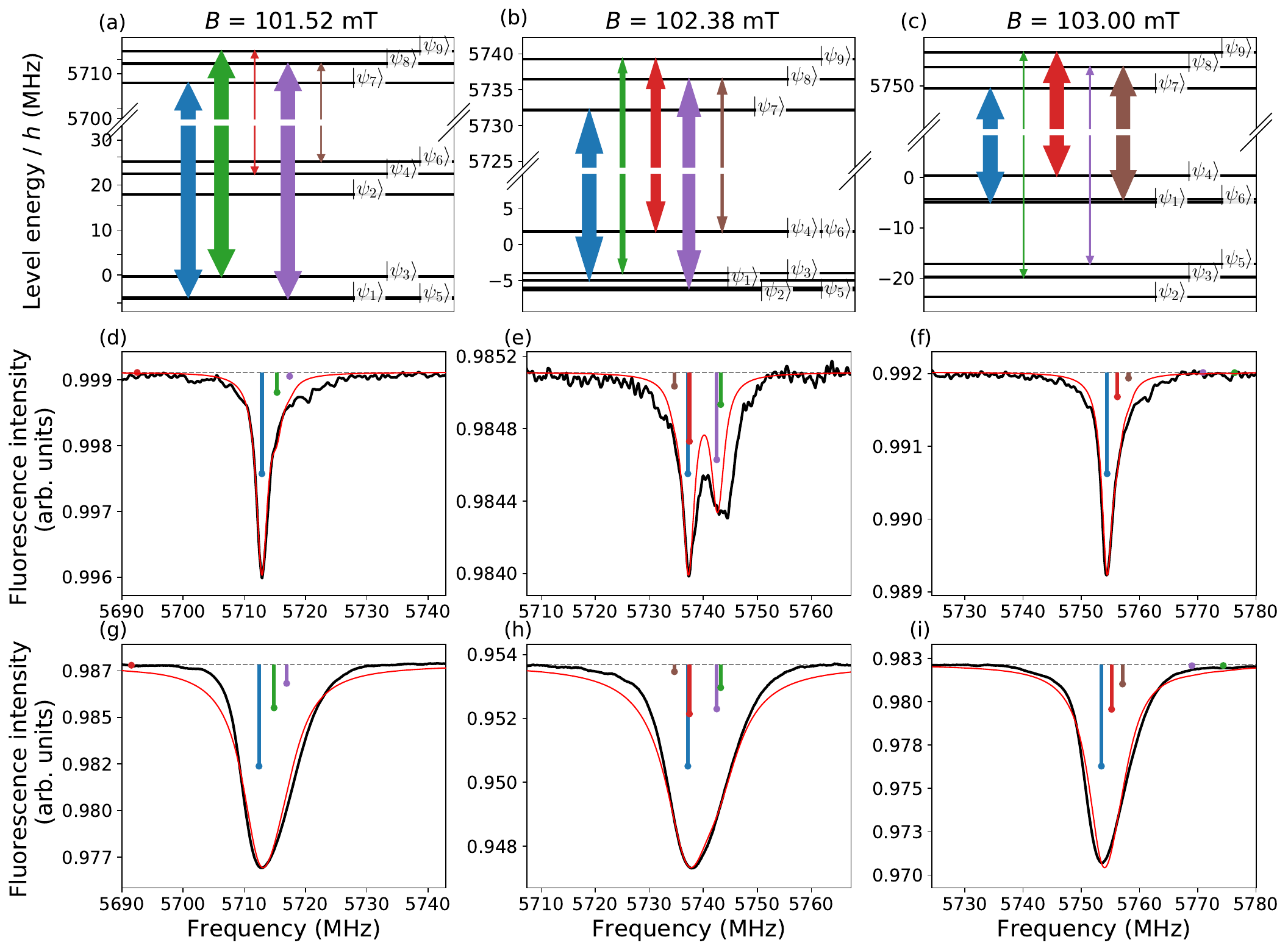}\
  \end{center}
  \caption{ODMR signals at high microwave field frequencies. The top row (a)--(c) shows the transitions between respective levels [Eqn. \eqref{eigenfunctions}]. The transition strength is indicated by the arrow width. In the middle row (d)--(f), the black lines show the experimental signals, and the red lines in (d)--(f) show the results of the theoretical calculations with the parameters from the fitting procedure described in the text for the low-density sample. The bottom row (g)--(i) shows the corresponding results for the high-density sample. The vertical bars in (d)--(i) correspond to the transitions depicted by the arrows in (a)--(c) of the same color, and their length determines the contribution to the overall lineshape of that transition, which is proportional to the product of the level population and the transition strength.}
  \label{high_fr}
\end{figure*}

Figure~\ref{high_fr} depicts ODMR signals for transitions in the frequency range 5.6--5.9~GHz, where the MW field is resonant with transitions from the mixed  $m_S=0$ and $m_S=-1$ hyperfine levels, to the $m_S=+1$ hyperfine levels [see Fig.~\ref{fig_gross}~(b)]. Experimentally measured signals are depicted together with curves obtained from a model calculation with some parameters obtained from a fitting procedure as explained below. Figures~\ref{high_fr}~(a)--(c) depict magnetic sublevels at a given magnetic field and indicate the allowed microwave transitions as arrows. The wave functions $|\psi_1 \rangle$--$|\psi_9 \rangle$ are given in Eq.~\eqref{eigenfunctions}. 
The middle row [Fig.~\ref{high_fr}~(d)--(f)] shows the experimental signals for the low-density sample, and the bottom row [Fig.~\ref{high_fr}~(g)--(i)] shows the signals for the high-density sample.

We used a parameter-optimization procedure based on a $\chi^2$ test to determine the contribution of each transition in Fig.~\ref{high_fr}~(a)--(c) to the overall lineshapes in Fig.~\ref{high_fr}~(d)--(i). The reduced $\chi^2$ value is defined as $\chi^2=(1/N)\Sigma [(d_i-f_i)/\sigma_i]^2$, where $N$ are the number of points, the $d_i$ are the measured data points, the $f_i$ are the results of the model, and the $\sigma_i$ are the mean square errors on the data points, which we set to unity here. To illustrate the procedure, let us consider Fig.~\ref{high_fr}~(d). This signal was recorded at a magnetic field value far away from the GSLAC. As a result, the mixing of the sublevels is insignificant, and a contribution to the signal is expected only from the following three transitions: $\vert 0,1 \rangle \longrightarrow \vert 1,1 \rangle$ (blue), $\vert 0,-1 \rangle \longrightarrow \vert 1,-1 \rangle$ (purple), and $\vert 0,0 \rangle \longrightarrow \vert 1,0 \rangle$ (green). In these transitions the nuclear spin projection $m_I$ does not change. We assumed that each transition has a Lorentzian lineshape centered at its respective transition frequency, which follows from the differences in level energies in Eq.~\eqref{energies}. The transition strengths for these three transitions are equal, which is indicated by the fact that all three arrows have the same width. Nevertheless, the relative contributions (peak amplitudes) of each transition may differ because of differences in the populations of the three ground states involved: $\vert 0, +1 \rangle$, $\vert 0, 0 \rangle$, and $\vert 0, -1 \rangle$, corresponding to the nuclear spin polarization of $^{14}\mathrm{N}$.

Next we attempted to find the 
spin temperature $\beta$ and magnetic field value $B$  
that minimized the reduced $\chi^2$ value for the hypothesis that our theoretical model with these parameters described the measured data. Far away ($>\pm 0.5$ mT) from the GSLAC the magnetic field value $B$ in the fit was allowed to vary over a small range since the position of the ODMR peak depends not only on $B$, but also on the nuclear spin polarization, which affects the contributions of transitions from the different nuclear spin components and thus can shift the ODMR peak position. Aside from the immediate vicinity of the GSLAC, the plot of ODMR peak position versus magnet coil current showed that the polarization does not  change very much with magnetic field over the range of 101.0 mT to 103.5 mT. Indeed, the fitted peak positions at each point produced a straight line as a function of current in the magnet coil that could be extrapolated through the region near the GSLAC point. In this way, a calibration curve for $B$ as a function of the magnet current was obtained, which was used to obtain the magnetic field value in the region near the GSLAC, where only the spin temperature was allowed to vary.

For each possible set of parameters $\beta$ and $B$, we calculated the corresponding populations of the eigenstates in~\eqref{eigenfunctions}.  We obtained the transition strengths from the calculated eigenvalues of these states. The amplitude of each transition peak is calculated according to \eqref{eq12}, the $m_s=+1$ states are assumed to be ``empty''. Then we used the SciPy optimize function~\cite{SciPyOpt} to determine the widths of the Lorentzians corresponding to each of the peaks, each one of which corresponds to a component of the hyperfine transition. This step was important because the width of the Lorentzians for our sample is around 1~MHz, and the peaks of nearby hyperfine components partially overlap. We assumed that all hyperfine components at a particular field strength had the same width. At the GSLAC the width of the Lorentzians increases due to increased relaxation rates arising from increased interaction between the hyperfine levels~\cite{Wood1}.

We calculated the reduced $\chi^2$ value using this set of parameters. We repeated the procedure for the next set of parameters and stored those that yielded the smallest reduced $\chi^2$ value. The peak amplitudes thus obtained are shown as the length of the colored bars in Fig.~\ref{high_fr}~(d). The color of each bar corresponds to the color of the arrow that represents the corresponding transition in Fig.~\ref{high_fr}~(a). We proceeded in a similar fashion for all subfigures. Near the GSLAC, there are more possible transitions that must be considered as a result of hyperfine level mixing [which follows from Eqn. \eqref{eigenfunctions}--\eqref{eigenfunctionsb}]. The number of possible transitions and their relative transition strengths are indicated by the number of arrows in Fig.~\ref{high_fr}~(b) and their widths.

\begin{figure*}%[tb]
  \begin{center}
    \includegraphics[width=\textwidth]{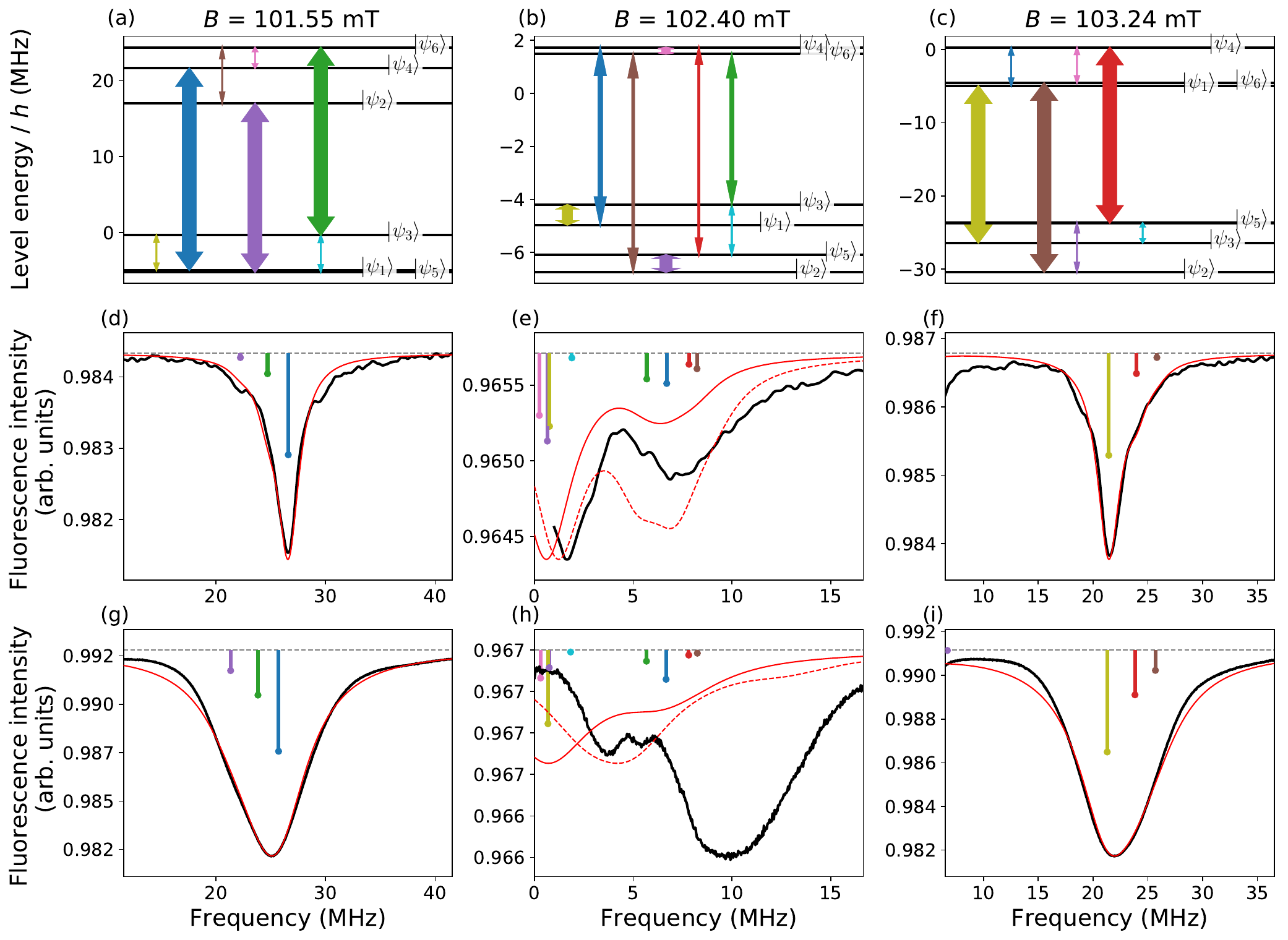}
  \end{center}
  \caption{ODMR signals at low microwave field frequencies. The top row (a)--(c) shows the transitions between respective levels [Eqn. \eqref{eigenfunctions}]. The transition strength is indicated by the arrow width.
  In the middle row (d)--(f), the black lines show the experimental signals and the red lines in (d)--(f) show the results of the theoretical calculations with the parameters from the fitting procedure described in the text for the low-density sample. The bottom row (g)--(i) shows the corresponding results for the high-density sample. The dashed red lines in (e) and (h) show the calculated signal for an angle between the NV axis and the magnetic field $\mathbf{B}$ of $\theta= 0.015^{\circ}$. In (d), (f), (g), and (i) there is no noticeable difference between calculated signals for $\theta= 0^{\circ}$ and $\theta= 0.015^{\circ}$}
  \label{low_fr}
\end{figure*}

\subsection{ODMR signals for the \texorpdfstring{$\vert m_S=0 \rangle \longrightarrow \vert m_S=-1 \rangle$}{|mS=0>-->mS=-1} transition}\label{sec:ODMR_lf}
We also measured ODMR signals for the $\vert m_S=0 \rangle \longrightarrow \vert m_S=-1 \rangle$ transition within $\pm 1.5$~mT of the GSLAC, which corresponds to microwave frequencies below 40~MHz, and some results are shown in Fig.~\ref{low_fr}. Experimentally measured signals are plotted together with signals from a model calculation using parameters that were obtained in a similar way as in Sec.~\ref{sec:ODMR_hf}. The top row (a)--(c) shows the magnetic-sublevel structure in a particular magnetic field with the allowed transitions depicted by arrows whose width indicates the relative transition strength. The middle row (d)--(f) shows the measured signals for the low-density sample, and the bottom row (g)--(i) shows the corresponding signals for the high-density sample. Again, above and below the GSLAC, the signals consist of three components, which correspond to the allowed transitions between hyperfine levels that are weakly mixed in the magnetic field, but at the GSLAC, there is strong mixing and more transitions must be taken into account, as indicated by the number of arrows in Fig.~\ref{low_fr}~(b). Above and below the GSLAC, the agreement between measured and calculated curves is quite good. However, right near the GSLAC there are some discrepancies in the amplitudes of the peaks. These discrepancies are particularly significant in the high-density sample, for which the model essentially fails at the GSLAC. Possible reasons for the discrepancies might be inhomogeneities in the microwave power, in the diamond crystal lattice or in the magnetic field, or interactions with other nearby spins, such as P1 centers or $^{13}$C nuclei. Interactions with nearby spins and unknown defects might be one more reason for the failure of the model at the GSLAC for the high-density sample. We consider the effect of $^{13}$C nuclei in Sec.~\ref{sec:C13}.

\begin{figure*}%[tb]
 \begin{center}
\includegraphics[width=\textwidth]{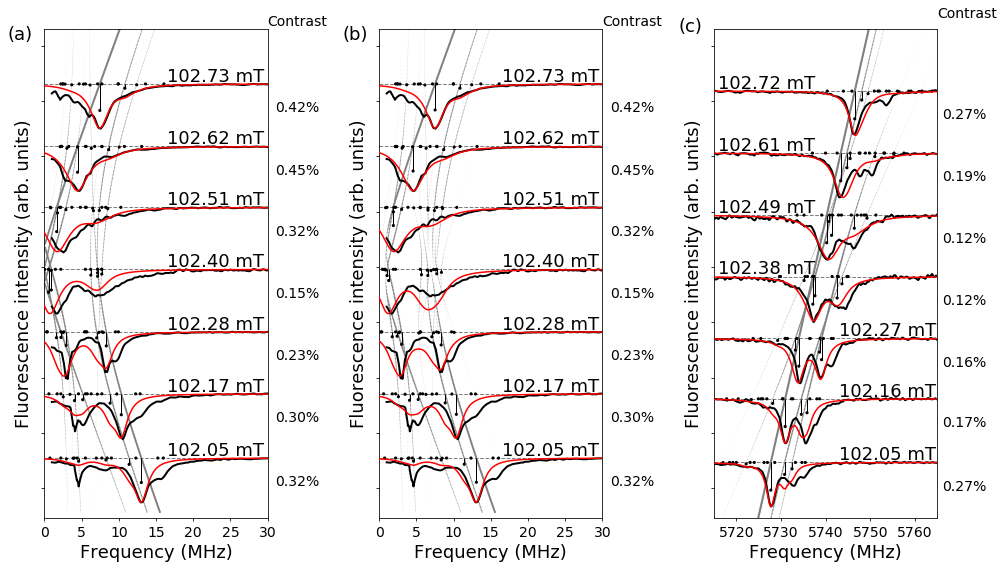}
\end{center}
\caption{Experimental signals (black) obtained on the low-density sample with theoretical calculations (red) for ground state $m_S = 0 \longrightarrow m_S = -1$ microwave transitions for different magnetic-field values. Experimental signal with the calculated signal for an angle between the NV axis and the magnetic field $\mathbf{B}$ of (a) $\theta= 0^{\circ}$ and (b) $\theta= 0.015^{\circ}$ (transverse magnetic field $0.025$~mT). (c) Experimental signal with the calculated signal at $\theta= 0^{\circ}$ for ground state $m_S = 0 \longrightarrow m_S = +1$ microwave transitions for different magnetic-field values. Bars with black dots on one end are placed at the values of the transition frequencies for a specific magnetic-field value, and their length represents the calculated transition probability. For better readability, signals are arranged in order of descending magnetic field, and each curve is normalized separately with its relative intensity depicted at the right side of the graph (see Fig.~\ref{ODMR_contrast} for details). The grey lines show how the energy and the intensity of the transitions change in the magnetic field.}
\label{zero_and_0075}
\end{figure*}

Figures~\ref{zero_and_0075}(a) and (b) show in more detail ODMR signals measured near the GSLAC for the $|m_S=0 \rangle \longrightarrow |m_S=-1 \rangle$ transition, and Fig.~\ref{zero_and_0075}(c) shows signals measured for the $|m_S=0 \rangle \longrightarrow |m_S=+1 \rangle$ transition in the low-density sample. The black curve shows the experimentally measured signals, while the red curve represents the result of the theoretical model calculation with parameters obtained by the same fitting procedure as described in connection with Figs.~\ref{high_fr} and~\ref{low_fr}. The percentages to the right of each frame show the actual ODMR contrast measured at that magnetic field value as given in Fig.~\ref{ODMR_contrast}. The curves plotted here are normalized, although the curve with a contrast of 0.45\% has in reality twice the amplitude as the curve with a contrast of 0.23\%.
Again, the nuclear spin populations and peak widths within $\pm 0.5$~mT of the GSLAC were taken from the values obtained for the high-frequency case at the same magnetic field value. Everywhere else the parameter optimization procedure was used as described in Sec.~\ref{sec:ODMR_hf}. The experimental data in Fig.~\ref{zero_and_0075}(a) and Fig.~\ref{zero_and_0075}(b) are identical, but for the calculated curves, the NV axis and the magnetic field vector were assumed to be parallel in the former case, whereas in the latter, an angle $\theta=0.015$ degrees between the NV axis and the magnetic field direction was assumed. This angle was found to give the best overall agreement between the experimentally measured values and the curves obtained from the model calculations with parameter fitting. In fact, it is difficult to align the magnetic field perfectly with the NV axis in the experiment. The signals in Fig.~\ref{high_fr} are relatively insensitive to small misalignment angles, which is why they were not plotted.

\begin{figure*}%[tb]
 \begin{center}
\includegraphics[width=\textwidth]{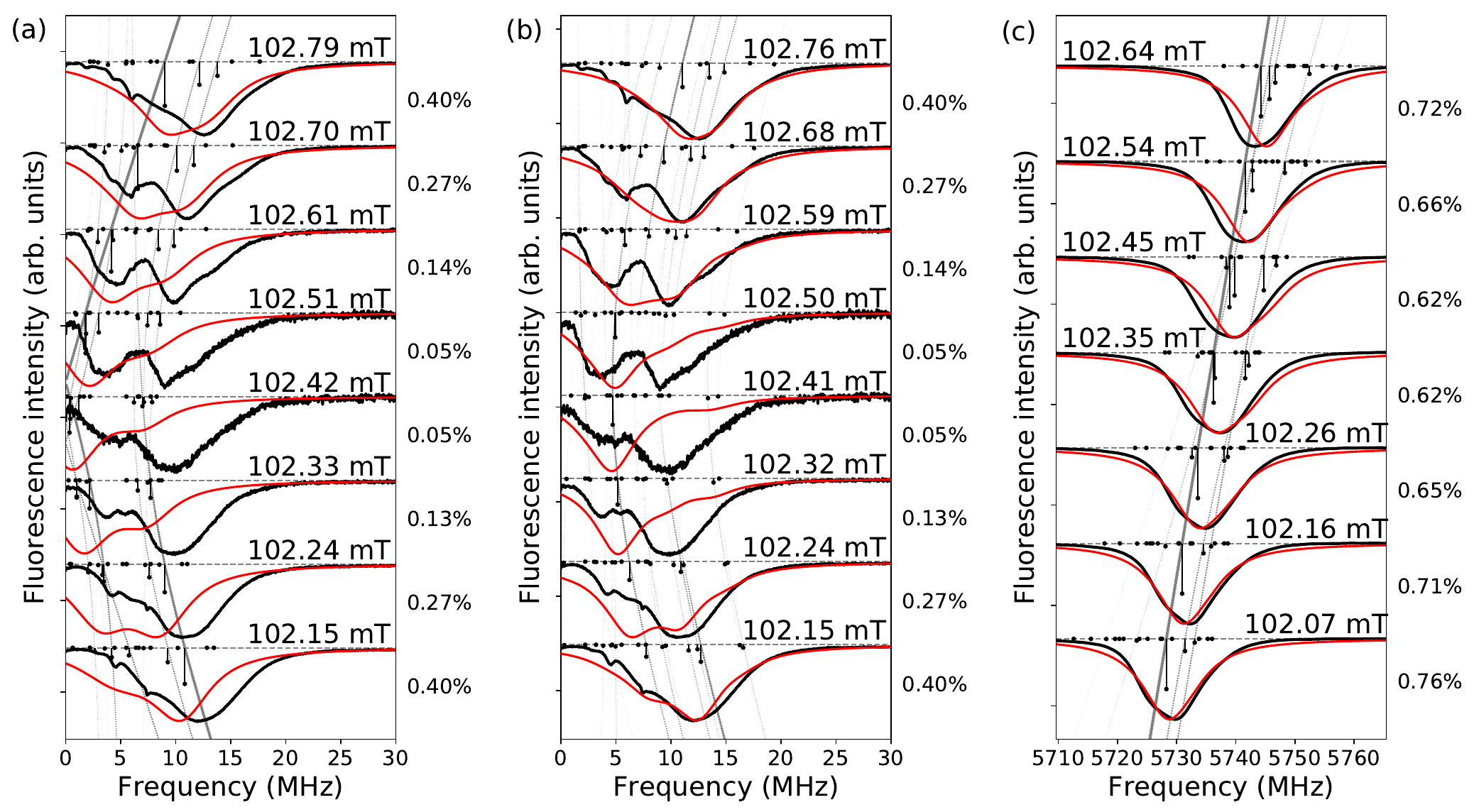}
\end{center}
\caption{Experimental signals from the high-density sample (black) with theoretical calculations (red) for ground state $m_S = 0 \longrightarrow m_S = -1$ microwave transitions for different magnetic-field values. Experimental signal \
with the calculated signal for an angle between the NV axis and the magnetic field $\mathbf{B}$ of (a) $\theta= 0^{\circ}$ and (b) $\theta= 0.1^{\circ}$ (transverse magnetic field $0.18$~mT). (c) Experimental signal with the calculated signal at $\theta= 0^{\circ}$ for ground state $m_S = 0 \longrightarrow m_S = +1$ microwave transitions for different magnetic-field values. Bars with black dots on one end are placed at the values of the transition frequencies for a specific magnetic-field value, and their length represents the calculated transition probability. For better readability, signals are arranged in order of descending magnetic field, and each curve is normalized separately with its relative intensity depicted at the right side of the graph (see Fig.~\ref{ODMR_contrast} for details). The grey lines show how the energy and the intensity of the transitions change in the magnetic field.}
\label{zero_and_0k1_Riga}
\end{figure*}

Similar curves are presented for the high-density sample in Figure~\ref{zero_and_0k1_Riga}. In this case, the angle used in Fig.~\ref{zero_and_0k1_Riga}(b) was $\theta=0.1^{\circ}$, which corresponds to a transverse magnetic field of 0.18 mT.

In all these experimentally measured signals, the overall peak intensities were normalized separately for each magnetic field value, because at the GSLAC, the contrast of the signals decreased strongly, as shown in Fig.~\ref{zero_and_0075}. The decrease in contrast near the GSLAC is caused by energy level mixing, which redistributes the population between the $\vert m_S=0 \rangle$ and the $\vert m_S=-1 \rangle$ levels. As a consequence, the $T_1$ time is drastically reduced, and so there are fewer NV centers in the ground state available for MW transitions. The vertical bars depict calculated ODMR peak positions and relative intensities. It can be seen that below and above the GSLAC the measured and calculated signals agree rather well. However, right at the GSLAC the agreement is not as good, and in particular, the model fails for the high-density sample at low frequencies in a range of $\pm 0.25$ mT of the GSLAC. The high-density sample may present additional defects that are not accounted for in the model or additional interactions among defects.

\begin{figure}%[tb]
  \begin{center}
    \includegraphics[width=0.45\textwidth]{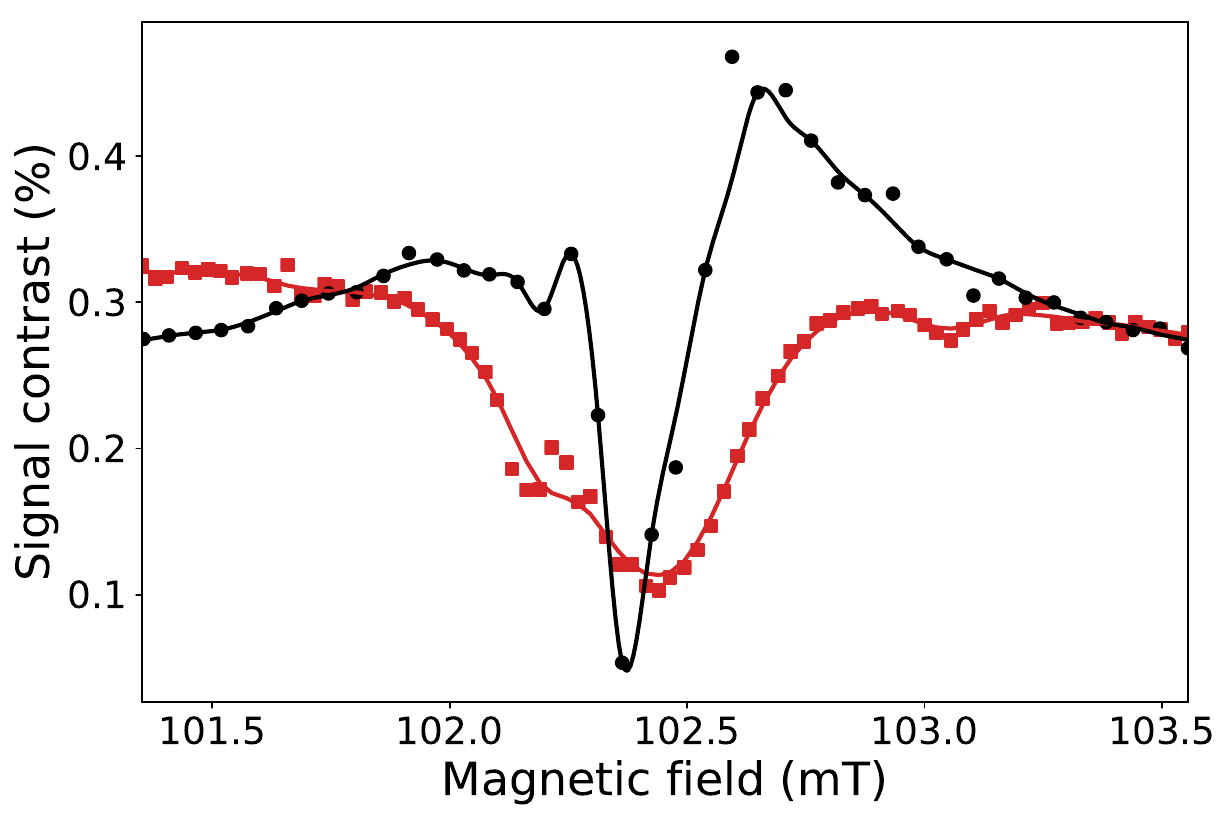}
  \end{center}
  \caption{ODMR signal contrast for different magnetic field values for the low-density sample. The line is drawn using locally weighted scatter plot smoothing. The image shows how the ODMR signal contrast changes close to the GSLAC region, black dots $m_S = 0 \longrightarrow m_S=-1$ transition (Fig. \ref{zero_and_0075}(a) and (b)), red squares $m_S = 0 \longrightarrow m_S=+1$ transition (Fig. \ref{zero_and_0075}(c)).}
  \label{ODMR_contrast}
\end{figure}

\subsection{Nuclear spin polarization} \label{sec:polarization}
The peak amplitudes from the fit in Fig.~\ref{low_fr}(d)--(f), Fig.~\ref{zero_and_0075}, and Fig.~\ref{high_fr}~(d)--(f) contain information about the relative populations of the ground-state hyperfine levels and thus the nuclear spin polarization~\cite{He1,Fuc1,Jac1,Sme1,Gal1,Sme2,Dre1,Bud3,Iva1}. Nuclear spin polarization arises from an interplay of optical pumping and sublevel mixing. For example, Fig.~\ref{fig_gross}~(c) shows how the $\vert m_S=0,m_I=-1 \rangle$ sublevel is mixed with the $\vert m_S=-1,m_I=0 \rangle$ sublevel near the GSLAC, where they are nearly degenerate. Optical pumping in this situation tends to move population from the $|m_S=0, m_I=-1 \rangle$ sublevel to the $|m_S=0, m_I=0 \rangle$ sublevel via the $|m_S=-1, m_I=0 \rangle$ sublevel. In a similar way, because of the mixing between $|m_S=0, m_I=0 \rangle$ sublevel and the $|m_S=-1, m_I=+1 \rangle$ sublevel, population moves from the $|m_S=0, m_I=0 \rangle$ sublevel to the $|m_S=0, m_I=+1  \rangle$ sublevel via the $|m_S=-1, m_I=+1 \rangle$ sublevel. However, the $|m_S=0, m_I=+1 \rangle$ sublevel is not mixed with any sublevel, and so the population accumulates in this state.

Besides spin temperature, another way to characterize the nuclear spin polarization is based on multipole expansion~\cite{ABR_2010}, in which the rank-zero multipole moment (monopole $\rho^0_0$) corresponds to population, the rank-one moment (dipole moment $\rho^1_0$), to orientation, and the second rank moment (quadrupole moment $\rho^2_0$), to alignment. In absence of processes that create coherences between different spin components in our system, only longitudinal (along the magnetic field direction) $\rho^1_0$ and $\rho^2_0$ spin polarization components are created.

Based on the component intensities, orientation would be calculated as
\begin{equation}
P^{1}_0 = \frac{\rho^1_0}{\rho^0_0} = \sqrt{\frac{3}{2}}\frac{n_{01}-n_{0-1}}
{n_{01}+n_{00}+n_{0-1}},\label{orientation}
\end{equation}
where $n_{m_S m_I}$ corresponds to the integral under the calculated curve that makes up all transitions from level $\left\vert m_S m_I\right\rangle$ (\ref{eigenfunctions}), and we assume that the $m_S=+1$ hyperfine levels are ``empty".

Alignment can be calculated in a similar way as
\begin{equation}
P^{2}_0 = \frac{\rho^2_0}{\rho^0_0} = 
\sqrt{\frac{1}{2}}\frac{n_{01}+n_{0-1} - 2 n_{00}}
{n_{01}+n_{00}+n_{0-1}}.\label{alignment}
\end{equation}

\begin{figure}%[tb]
  \begin{center}
    \includegraphics[width=0.45\textwidth]{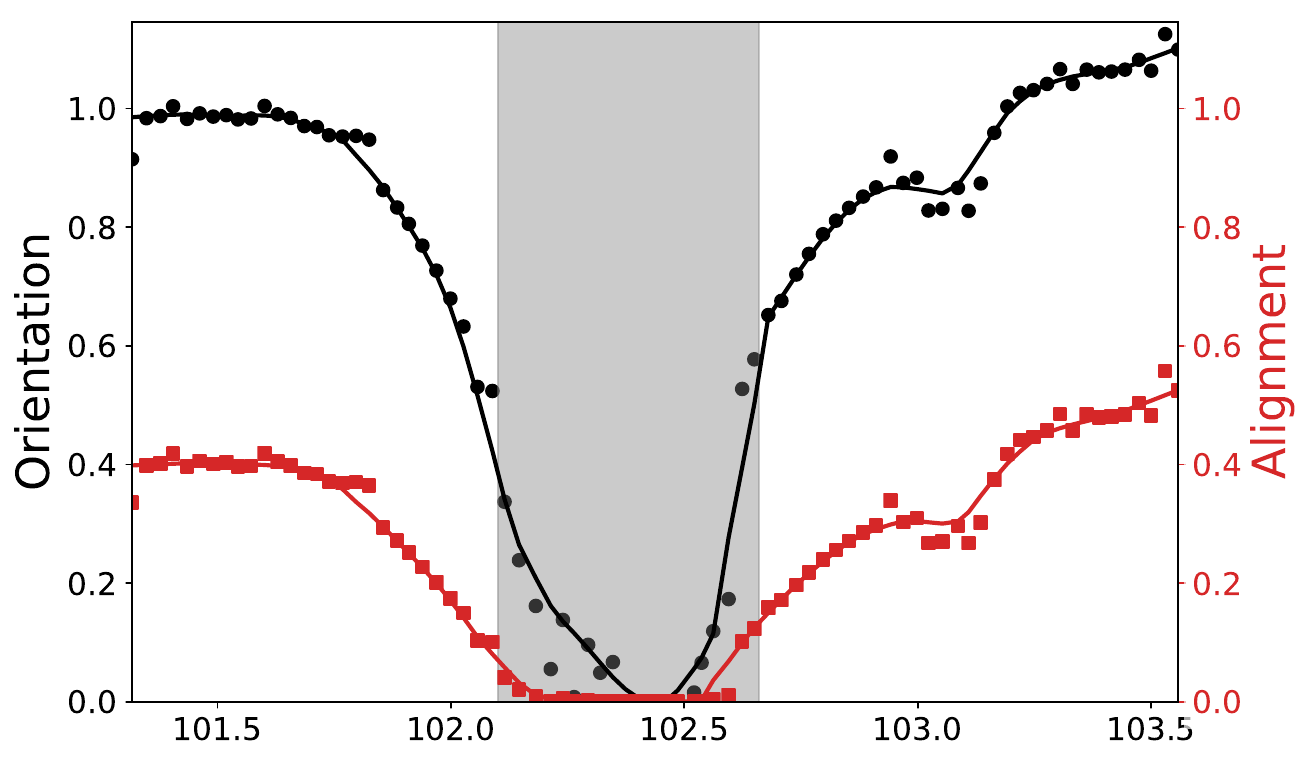}
  \end{center}
  \caption{Nuclear $^{14}$N spin orientation (black dots) and alignment (red squares) obtained for the ground state $m_S = 0 \longrightarrow m_S = +1$ transition from the integral under the fitted transition peaks for the low-density sample. The solid lines are drawn using locally weighted scatter plot smoothing. The gray, shaded region corresponds to the magnetic field range in which we cannot be certain about the polarization value because an additional peak poorly described by the model interferes with the optimization procedure.
}
  \label{fig:orientation}
\end{figure}

Figure~\ref{fig:orientation} shows the degree of spin polarization as a function of magnetic field obtained for the low-density sample. The high-density sample is not used, since the model fails near the GSLAC. The populations of the nuclear spin components were obtained from the peak amplitudes obtained by the parameter fitting procedure described in Sec.~\ref{sec:ODMR_hf}, and then the orientation and alignment were calculated using Eqs.~\eqref{orientation} and~\eqref{alignment}, respectively. The data for this procedure are shown in Fig.~\ref{high_fr}, supplemented by similar measurements at many more magnetic-field values. The population of each nuclear spin component is proportional to the integral under the corresponding calculated curve divided by the transition strength. The region at the center shaded in gray in Fig.~\ref{fig:orientation} indicates the magnetic field values for which an additional peak poorly described by the model [see, for example, Fig.~\ref{high_fr}(e)] complicates the optimization procedure, and makes it difficult to guarantee that the polarization value obtained from this procedure fully describes the true polarization. The polarization values extracted by our method are plotted in this region as well, but they should be taken with caution. 
Both figures show nuclear spin polarization near the GSLAC with a minimum at the GSLAC position. 
In principle, the polarization should vanish far away from the GSLAC where there is no hyperfine level mixing~\cite{Iva1}. However, we did not observe this behavior over the measured range of magnetic field values.

\subsection{Influence of \texorpdfstring{$^{13}$}{13}C Nuclei} \label{sec:C13}
We now consider the interaction of our system with a collection of nearby $^{13}{\mathrm C}$ nuclei labeled by the index $j$. The interaction between the NV center and these nuclei is described by the Hamiltonian\cite{Nizovtsev2010}:
\begin{equation}
    H_{NV+13C} = \Sigma_j \left( \mathbf{\hat{S}} \cdot A'_{C13,j} \cdot \mathbf{\hat{J}}_j + \hat{H}_{13C_j} \right) \; ,
    \label{Hamiltonian_13C}
\end{equation}
where 
$\mathbf{\hat{J}}_j$ labels the nuclear spin of the $j$-th $^{13}{\mathrm C}$ nucleus and $\hat{H}_{13C_j}=\gamma_{13C} \mathbf{B} \cdot \mathbf{\hat J}_j$ is the Hamiltonian corresponding to the $j$-th $^{13}{\mathrm C}$ nucleus. 
The tensor $\hat{A}_{C13,j}$ has the same form as $\hat{A}$ in eq.~\eqref{AN14}. For the case of a $^{13}{\mathrm C}$ nucleus in one of the three lattice positions next to the vacancy, $\hat{A}_{C13,j=1}$, $A_{\perp}=A_{xx}=A_{yy} =121.1$~MHz and $A_{\parallel} = A_{zz}=199.21$~MHz~\cite{Gal1}.  
We follow a procedure similar to the one outlined by Nizovtsev \textit{et al.}~\cite{Nizovtsev2010} to 
rotate this tensor from the principal coordinate axes of the carbon nucleus to the coordinate system of the NV center with the $Z$-axis parallel to the [111] crystal direction. The rotation takes the form of $\hat{A}'=\hat{U} \hat{A} \hat{U}^T$, where $\hat{U}$ is the rotation matrix about the $x$-axis by an angle $\cos(zZ)$ that rotates the $z$-axis in the frame of the carbon nucleus into the frame of the [111] crystal direction.

The principal values of the tensor $\hat{A}_{C13,j}$ that corresponds to other lattice positions besides the nearest neighbors were taken from density functional theory (DFT) calculations performed by Gali \textit{et al.}~\cite{Gali2008}. The lattice positions can be classified according to families of lattice points that have the same tensor values~\cite{Sme2} (see Fig.~4 and Table~1 in that publication, which also gives the multiplicities of each family.) The value of $\vert \cos(zZ) \vert$ of the angle between the $z$-axis of the carbon nucleus's principal axis and the NV axis were taken from the results of another DFT calculation performed by Nizovtsev et al.~\cite{Nizovtsev2014}. We took into account the nearest neighbors of the vacancy site as well as families A through H, which corresponds to 39 lattice sites. Other lattice sites have significantly less influence.    

Next we use a Monte Carlo method to average scenarios where $^{13}$C atoms are located in different lattice sites. For each iteration, we loop through the 39 lattice sites, each of which has a 1.1\% probability of hosting a $^{13}$C atom. Those that contain a $^{13}$C atom are added to the Hamiltonian in ~\eqref{Hamiltonian_13C} with the values of the hyperfine tensor as described in the previous paragraph. 
The number of energy levels of the system $N_{levels}$ follows from the eigenvalues and eigenvectors of Eq.~\eqref{eq1}, and it depends on the number $N_{C13}$ of $^{13}{\mathrm C}$ atoms as $N_{levels}=3\times3\times 2^{N_{C13}}$. The spectrum generated depends on the locations of the $^{13}$C atoms, which can differ from iteration to iteration. Thus, several hundred iterations are averaged to take into account that each NV center contributing to our signal has a different arrangement of $^{13}$C atoms in its vicinity.

\begin{figure}%[tb]
 \begin{center}
\includegraphics[width=0.45\textwidth]{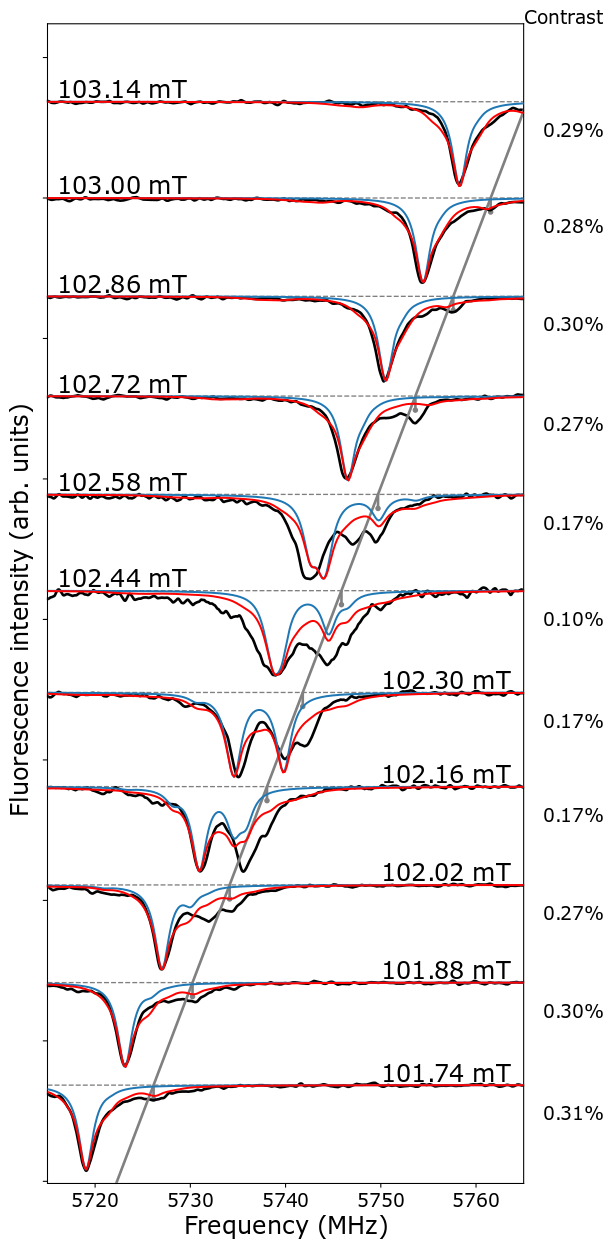}
\end{center}
\caption{Experimental signals (black) for the low-density sample with theoretical calculations including the effects of $^{13}\mathrm{C}$(red) [see \eqref{Hamiltonian_13C}] and without the effects of $^{13}\mathrm{C}$ (blue) for the ground state $m_S = 0 \longrightarrow m_S = +1$ microwave transitions for different magnetic-field values. The grey line tracks the position of the nominal $\vert 0,1\rangle \longrightarrow \vert 1,0\rangle$ transition.}
\label{c13_high}
\end{figure}

\begin{figure}%[tb]
 \begin{center}
\includegraphics[width=0.45\textwidth]{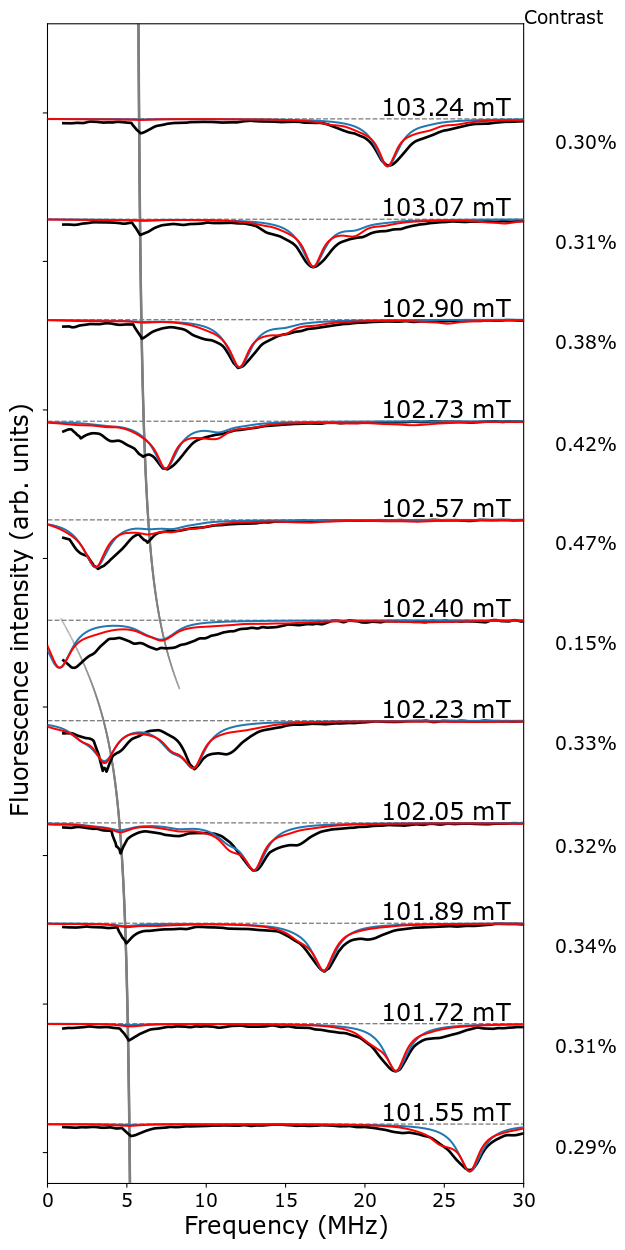}
\end{center}
\caption{Experimental signals (black) for the low-density sample with theoretical calculations including the effects of $^{13}$C(red) [see \eqref{Hamiltonian_13C}] and without the effects of $^{13}$C (blue) for the ground state $m_S = 0 \longrightarrow m_S = -1$ microwave transitions for different magnetic-field values. The grey line tracks the position of the nominal $\vert 0,1\rangle \longrightarrow \vert 0,0\rangle$ transition.}
\label{c13_low}
\end{figure}

The results of the model calculations with $^{13}$C are shown in Fig.~\ref{c13_high} for the $m_S=0 \longrightarrow m_S=+1$ transition and in Fig.~\ref{c13_low} for the $m_S=0 \longrightarrow m_S=-1$ transition. In both figures, the blue curves show the calculations without the $^{13}$C interaction, whereas the red curves include the interaction with nearby $^{13}$C nuclei. One can see that the inclusion of the $^{13}$C interaction slightly improves the agreement with the calculations. Only the low-density sample was used for this comparison, since the model was more successful for this sample. Nevertheless, discrepancies remain, which might require the inclusion of more terms in the Hamiltonian, such as nearby P1 centers, which is outside the scope of this study.

The addition of $^{13}$C nuclei in the Hamiltonian [see Eq.~\eqref{eq1}] allows for additional transitions to appear in the calculated signals. These new transitions are most apparent in describing the shoulder to the right of the main peak in the top three and bottom two panels of Fig.~\ref{c13_low}, as well as to the left of the main peak in the top two panels. Without the inclusion of $^{13}\mathrm{C}$ nuclei in the model, these shoulders cannot be described. A similar effect can be observed in Fig.~\ref{c13_high}, but not as dramatically.
The influence of $^{13}\mathrm{C}$ nuclei allows otherwise forbidden transitions because the majority of them are off-axis, which effectively changes the angular momentum selection rules~\cite{Wang2}.
The gray curve in Fig.~\ref{c13_high} corresponds to the $\vert 0,1\rangle \longrightarrow \vert 1,0\rangle$ transition, which generally appears in the experimentally measured signals, but is not so pronounced in either of the theoretical models, although the model that includes $^{13}\mathrm{C}$ hints at it far away from the GSLAC position. 

In some cases there remain significant discrepancies between the model and the experiment, even far from the GSLAC position. For example, the gray curve in Fig.~\ref{c13_low} tracks the nominal $\vert 0,1\rangle \longrightarrow \vert 0,0\rangle$ transition. This transition corresponds to a strong peak in the experimentally measured signals; this peak appears also in the calculation, but is so weak that it cannot be distinguished at the magnification shown in the figure. 

We conclude that the influence of $^{13}\mathrm{C}$ reveals important aspects of the underlying physics and influences the strengths of the transitions that can take place, but fails to explain other significant aspects of the measured signals. Perhaps more detailed calculations that take into account the spin dynamics could reproduce the ODMR spectra more accurately.

We also note that there is a peak that appears in a number of the figures just to the right of the main peaks that cannot be explained by the model. Interestingly, this peak tracks the $\vert 0,1\rangle \longrightarrow \vert 1,0\rangle$ transition as is shown by the gray line in Fig.~\ref{c13_high}. This transition should not occur according to Eq.~\eqref{transition_strengths}. However, the measurements suggest that some interaction (possibly off-axis $^{13}\mathrm{C}$ as well) causes this transition to occur. Thus,
a search for this interaction should also be considered in future theoretical work.

\section{Conclusions}
In this work we have studied microwave-induced transitions between the hyperfine components of the $^{3}A_2$ ground-state sublevels of the NV center in diamond using the ODMR technique in two diamond samples with a nitrogen concentration of 1~ppm and 200~ppm. We have developed a straightforward theoretical model to describe these ODMR signals. The model describes the ODMR signals for magnetic field values in the vicinity of the GSLAC as well as away from it. Including the effects of nearby $^{13}\mathrm{C}$ nuclei significantly improves the agreement between the model and the experimentally measured signals. The theoretical model allows to track $\vert 0,1\rangle\longrightarrow\vert 0,0\rangle$ and $\vert 0,1\rangle\longrightarrow \vert 1,0\rangle$ transitions enabled by the hyperfine level mixing.  Within 0.5~mT of the GSLAC, the experimentally measured ODMR spectrum becomes rather complicated with some features that we have not been able to describe fully, although the general features are reproduced at least in the case of the low-density sample. In the case of the high-densitiy sample, the model fails at low frequency in the range of $\pm 0.25$~mT around the GSLAC position. Further investigation will be needed to find the interactions responsible for all of these features. Nevertheless, the model can be used to estimate the nuclear spin polarization of $^{14}\mathrm{N}$, which we have demonstrated in this study. We have performed ODMR measurements on the $\vert m_S=0\rangle \longrightarrow \vert m_S=-1\rangle$ transition as well as on the $\vert m_S=0\rangle \longrightarrow \vert m_S=+1\rangle$ transition. The latter transition is somewhat simpler, since the $\vert m_S=+1\rangle$ level is not involved in the hyperfine level mixing and anticrossing and thus serves as a useful cross-check to test the adequacy of the theoretical model. The ODMR technique has proven to be a useful tool for investigating how the hyperfine interaction influences the energy-level structure near the GSLAC and, with further improvements to the theoretical model, could shed more light on additional interactions and the process of nuclear spin polarization near the GSLAC. The results of this work will be used in the ongoing efforts to model and optimize NV-diamond based microwave-free sensors, in particular, magnetometers~\cite{Zheng1,Wickenbrock1}.

\section{Acknowledgements}
We thank Daniels Krimans for helpful contributions to the theoretical approach. The Riga group gratefully acknowledges the financial support from the M-ERA.NET project “Metrology at the Nanoscale with Diamonds” (MyND), from the Laserlab-Europe Project (EU-H2020 654148), and from the Base/Performance Funding Project Nr. AAP2016/B013, ZD2010/AZ27. A.~Berzins acknowledges support from the PostDoc Latvia Project Nr. 1.1.1.2/VIAA/1/16/024 "Two-way research of thin-films and NV centres in diamond crystal". 

The Mainz group acknowledges support by the German Federal Ministry of Education and Research (BMBF) within the Quantumtechnologien program (FKZ 13N14439) and the DFG through the DIP program (FO 703/2-1). H. Zheng acknowledges  support from the GRK Symmetry Breaking (DFG/GRK 1581) program.
This work was also supported by the EU FET OPEN Flagship Project ASTERIQS (action 820394) and the Cluster of Excellence Precision Physics, Fundamental Interactions, and Structure of Matter (PRISMA+ EXC2118/1) funded by the German Research Foundation (DFG) within the German Excellence Strategy (Project
ID 39083149).

\bibliography{Anticrossing}
\bibliographystyle{apsrev}

\end{document}